\documentclass[pra,twocolumn,nofootinbib]{revtex4}

\usepackage{graphicx}
\usepackage{dcolumn}
\usepackage{bm}
\usepackage{amssymb}
\usepackage{amsmath}
\usepackage{amsfonts}
\usepackage[]{graphicx}
\def\beq{\begin{eqnarray}}
\def\eeq{\end{eqnarray}}

\begin{document}
\bibliographystyle{prsty}

\title{In defense of Relational Quantum Mechanics: A note on `Qubits are not observers' }

\author{Aur\'elien Drezet}
\affiliation{Institut N\'eel, UPR 2940, CNRS-Universit\'e Joseph Fourier, 25, rue des Martyrs, 38000 Grenoble, France}

\date{\today}

\begin{abstract}
This is a short note to answer Brukner's objection [see arXiv:2107.03513] to Rovelli's theory and concerning the preferred basis problem.  
\end{abstract}
\maketitle

\indent The aim of the present note is to give a short reply  to the recent C.~Brukner paper\cite{Brukner} (see also \cite{Pienaar}) concerning the Relational Quantum Mechanics interpretation (RQM) proposed by C.~Rovelli~\cite{Rovelli1996,RovelliBook2021,Laudisa,RovelliFP}.  Recently, Di Biagio and Rovelli offerred a reply~\cite{Rovellian}. Here, the aim is to give an even shorter comment.\\
\indent  The main issue concerns the interpretation of the full wavefunction  $|\Psi_{OS}\rangle$ involving observer (O) and observed system (S).  In RQM the fundamental object  \underline{relatively to (O)} is not   $|\Psi_{OS}\rangle$ but the reduced density matrix
\begin{eqnarray}
\hat{\rho}^{(red.)}_S=\textrm{Tr}_O[\hat{\rho}_{OS}]=\textrm{Tr}_O[|\Psi_{OS}\rangle \langle \Psi_{OS}|].
\end{eqnarray} 
As it is well known $\hat{\rho}^{(red.)}_S$ is independent of the basis chosen to represent the degrees of freedom for (O).  This solves the dilemma discussed in \cite{Brukner,Pienaar} concerning the `preferred basis problem'.\\
\indent More precisely, considering an EPR state like 
\begin{eqnarray}
\frac{|\textrm{here}_S\rangle\otimes|\ddot{\smile}_O\rangle+|\textrm{there}_S\rangle_S\otimes|\ddot{\frown}_O\rangle}{\sqrt{2}}\label{eq1}
\end{eqnarray} where $|\textrm{here}_S\rangle,|\textrm{there}_S\rangle $ and $|\ddot{\smile}_O\rangle,|\ddot{\frown}_O\rangle$ are bases for (S) and (O) respectively.
 Eq.~\ref{eq1} can alternatively be written as  
\begin{eqnarray}
\frac{1}{\sqrt{2}}(\frac{|\textrm{here}_S\rangle+|\textrm{there}_S\rangle}{\sqrt{2}})\otimes(\frac{|\ddot{\smile}_O\rangle+|\ddot{\frown}_O\rangle}{\sqrt{2}})\nonumber\\
+\frac{1}{\sqrt{2}}(\frac{|\textrm{here}_S\rangle-|\textrm{there}_S\rangle}{\sqrt{2}})\otimes(\frac{|\ddot{\smile}_O\rangle-|\ddot{\frown}_O\rangle}{\sqrt{2}}).
\label{eq2}\end{eqnarray} As it is well known this leads to the preferred basis problem in quantum measurement theory~\cite{Peres,Zurek}.\\
\indent   Now, in RQM for (O) the fundamental quantity is $\hat{\rho}^{(red.)}_S$ [for (S) this is the symmetrically the matrix $\hat{\rho}^{(red.)}_O=\textrm{Tr}_S[\hat{\rho}_{OS}]$]. Therefore, instead of Eq.~\ref{eq1} or \ref{eq2} we must consider the matrix:
\begin{eqnarray}
\hat{\rho}^{(red.)}_S=\frac{1}{2}|\textrm{here}_S\rangle\langle \textrm{here}_S|+ \frac{1}{2}|\textrm{there}_S\rangle\langle \textrm{there}_S| \label{eq3}
\end{eqnarray} that is directly equivalent to \cite{Penrose} 
 \begin{eqnarray}
\hat{\rho}^{(red.)}_S=\frac{1}{2}|\textrm{\underline{here}}_S\rangle\langle \textrm{\underline{here}}_S|+ \frac{1}{2}|\textrm{\underline{there}}_S\rangle\langle \textrm{\underline{there}}_S| \label{eq3}
\end{eqnarray} with \begin{eqnarray}
|\textrm{\underline{here}}_S\rangle=\frac{|\textrm{here}_S\rangle+|\textrm{there}_S\rangle}{\sqrt{2}}\nonumber\\
|\textrm{\underline{there}}_S\rangle=\frac{|\textrm{here}_S\rangle-|\textrm{there}_S\rangle}{\sqrt{2}}.
\end{eqnarray}  The description is thus basis independent. Moreover, the observer (O) doesn't care in RQM about the basis $|\ddot{\smile}_O\rangle,|\ddot{\frown}_O\rangle$ or $\frac{|\ddot{\smile}_O\rangle\pm|\ddot{\frown}_O\rangle}{\sqrt{2}}$ which is not appearing in  $\hat{\rho}^{(red.)}_S$ and there is no paradox in RQM.\\
\indent We  mention en passant that this issue about density matrix in RQM is interesting to decipher the role of locality/nonlocality. Furthermore, some kind of preferred basis could in the end be introduced in RQM if we try to formulate it as an hidden variable theory à la de Broglie-Bohm (as the author did recently for himself):  But this is a different story.

\end{document}